\documentclass[onecolumn,showpacs]{revtex4}

\usepackage{graphicx}
\usepackage{dcolumn}
\usepackage{bm}

\input epsf

\begin{document}

\title{\Large Adiabatic Particle Production with Decaying $\Lambda$ and Anisotropic Universe}

\author{\bf Sisir Bhanja}
\author{\bf Subenoy Chakraborty}
\email{subenoyc@yahoo.co.in}
\author{\bf Ujjal Debnath}
\email{ujjaldebnath@yahoo.com}

\affiliation{Department of Mathematics, Jadavpur University,
Calcutta-32, India.\\\\\\}

\date{\today}

\begin{abstract}
Here we study an anisotropic model of the universe with constant
energy per particle. A decaying cosmological constant and particle
production in an adiabatic process are considered as the sources
for the entropy. The statefinder parameters $\{r,s\}$ are defined
and their behaviour are analyzed graphically in some cases.
\end{abstract}


\maketitle

The inclusion of cosmological constant into Einstein field
equation was done primarily by Einstein himself with a view to
model a static universe. Later he discarded the cosmological
constant (and commented [1] ``...... the biggest blunder of my
life''), on the basis of the observational data supporting the
expansion of the universe. Subsequently, it was reintroduced and
then omitted to match with the observational evidences.\\

Recently, cosmological models with a cosmological constant are
strong candidates to describe the dynamics of the universe. The
observational results [2, 3] from type Ia Supernova suggests that
our universe is accelerating. The increase of the expansion
velocity can be accounted for by introducing a repulsive force
which results a negative pressure in the energy-momentum tensor.
Cosmological constant (or term, time dependent) is a good
candidate to generate negative pressure and to account the vacuum
contribution to the energy-momentum tensor. Further, the present
small value of $\Lambda$ [4, 5] suggests that it is not a constant
but have a dynamical evolution. As universe expands, the effective
cosmological term evolves and decreases to its present value. This
is also supported strongly from observational point of view,
particularly, the anisotropy of the cosmic microwave background
radiation, the supernova experiments and estimated age of the
universe.\\

Moreover, the $\Lambda$-term can be considered as a type of dark
non-baryonic matter (or dark matter) in the sense that it is not
gravitationally clustered at all scales as the usual matter.
Lastly, it is to be noted that in the literature there are
cosmological models with particle production which are compatible
with experimental evidences.\\

In this work, using the adiabatic condition, the dissipative
pressure is considered due to decaying of a cosmological term and
particle production. The thermodynamical condition is necessary
for the conformity of the cosmic microwave background radiation
due to creation of photons.\\

In thermodynamics, the energy conservation equation is

\begin{equation}
nTd\sigma=d\rho-(\mu+T\sigma)dn
\end{equation}

where the chemical potential $\mu$ satisfies the Euler equation

\begin{equation}
\mu=\frac{(\rho+p)}{n}-T\sigma
\end{equation}

So eliminating $\mu$ between (1) and (2) we have

\begin{equation}
nT\dot{\sigma}=\dot{\rho}-\frac{(\rho+p)}{n}~\dot{n}
\end{equation}
(`~.~' stands for time derivative)\\

The number density `$n$' satisfies the continuity equation

\begin{equation}
\dot{n}+n\Theta=\Psi
\end{equation}

with $\Theta$, the expansion scalar and $\psi$, the source of
particles. Now the second law of thermodynamics can be written as

\begin{equation}
S^{\alpha}_{,\alpha}=n\dot{\sigma}+\sigma\dot{\Psi}
\end{equation}

with $S^{\alpha}=n\sigma u^{\alpha}$ as the entropy four vector.\\

Now, considering the energy-momentum tensor as

\begin{equation}
T^{\mu\nu}=(\rho+p_{_{T}})u^{\mu}u^{\nu}-p_{_{T}}g^{\mu\nu}+\Lambda
g^{\mu\nu}
\end{equation}

the energy conservation equation ($T^{\mu\nu}_{~;\nu}=0$) results

\begin{equation}
\dot{\rho}+(\rho+p_{_{T}})\Theta=-\frac{\dot{\Lambda}}{8\pi G}
\end{equation}

Here $\rho$ is the matter density, $p_{_{T}}$ is the sum of the
usual thermodynamical pressure $p$ and a dissipative pressure
$\Pi$ (i.e., $p_{_{T}}=p+\Pi$), $u^{\mu}$ is the four velocity
vector and $\Lambda=\Lambda(t)$, a varying cosmological
constant.\\

Now substituting $\dot{\sigma}$ and $\dot{n}$ from equation (3)
and (4) and using the energy conservation equation (7) we have
from equation (5), the expression for the entropy as

\begin{equation}
S^{\alpha}_{;~ \alpha}=-\frac{n}{T}\left[\frac{\dot{\Lambda}}{8\pi
G}+\frac{(\rho+p)}{n}~\Psi+\Pi\Theta \right]+\sigma\dot{\Psi}
\end{equation}

Considering the particle production process as an adiabatic
process (i.e., $\dot{\sigma}=0$) one writes

\begin{equation}
\frac{\dot{\Lambda}}{8\pi G}+\frac{(\rho+p)}{n}~\Psi+\Pi\Theta=0
\end{equation}

The line element for the Kantowski-Sachs (KS) model is given by

$$
ds^{2}=-dt^{2}+a^{2}(t)dr^{2}+b^{2}(t)d\Omega_{2}^{2}~,
$$

The Einstein field equations for the matter content of the form
(6) is given by

\begin{equation}
2\frac{\ddot{b}}{b}+\frac{\dot{b}^{2}}{b^{2}}
+\frac{k}{b^{2}}=-8\pi G(p+\Pi)+\Lambda
\end{equation}

\begin{equation}
\frac{\ddot{a}}{a}+\frac{\ddot{b}}{b}+\frac{\dot{a}}{a}\frac{\dot{b}}{b}
=-8\pi G(p+\Pi)+\Lambda
\end{equation}
and
\begin{equation}
\frac{\dot{b^{2}}}{b^{2}}
+2\frac{\dot{a}}{a}\frac{\dot{b}}{b}+\frac{k}{b^{2}}=8\pi G
\rho+\Lambda
\end{equation}

Assuming barotropic equation of state $p=\epsilon\rho$
($0<\epsilon<1$), one can eliminate $\rho$, $p$ and $\Pi$ using
the field equations (10)-(12) and equation (9) to obtain

\begin{eqnarray*}
2\frac{\ddot{b}}{b}+\left(\frac{\dot{b}^{2}}{b^{2}}+\frac{k}{b^{2}}\right)\left\{(1+\epsilon)
\left(1-\frac{\Psi}{n\Theta}\right)\right\}+2\frac{\dot{a}}{a}\frac{\dot{b}}{b}
\left\{(1+\epsilon) \left(1-\frac{\Psi}{n\Theta}\right)-1\right\}
\end{eqnarray*}
\begin{equation}
=\frac{\dot{\Lambda}}{\Theta}+\Lambda(1+\epsilon)
\left(1-\frac{\Psi}{n\Theta}\right)\hspace{-1.5in}
\end{equation}
with $\Theta=\frac{\dot{a}}{a}+2\frac{\dot{b}}{b}=3H$.\\

In the present model, particle production is possible due to
variation of $\Lambda$ and due to dissipative pressure. We now
assume, the contribution due to these terms are proportional
namely,

\begin{equation}
\frac{\dot{\Lambda}}{8\pi G}=\omega(\rho+p)\frac{\Psi}{n}
\end{equation}

with $\omega$ as the constant of proportionality. It is to be
noted that vanishing of $\Psi$ implies both $\Lambda$ and entropy
to be constant.\\

Combining equations (9) and (14), the dissipative pressure is
given by

\begin{equation}
\Pi=-(1+\omega)(\rho+p)\frac{\Psi}{n\Theta}
\end{equation}

Also taking into account relations (7), (9) and the field
equations, one gets

\begin{equation}
(\rho+p)\Theta\left(1-\frac{\Psi}{n\Theta}\right)=\frac{1}{8\pi
G}\left(\Lambda-3H^{2}\right)^{\dot{}}
\end{equation}

Now choosing $\Lambda=\alpha H^{2}$ as proposed by Viswakarma [6]
in recent past, we see that the particle source corresponding to
the $\Lambda$ term is

\begin{equation}
\Psi=\beta n \Theta
\end{equation}

where $\beta$ is a phenomenological constant [7].\\

Using these relations, (namely, (15)-(17)) in the field equations
to eliminate $\rho$, $p$ and $\Pi$, we have the differential
equation in the scale factors as

\begin{eqnarray*}
2\frac{\ddot{b}}{b}+\frac{\dot{b}^{2}}{b^{2}}(1+\epsilon)\{1-(1+\omega)\beta\}
-2\frac{\dot{a}}{a}\frac{\dot{b}}{b}[1-(1+\epsilon)\{1-(1+\omega)\beta\}]+\frac{k}{b^{2}}
(1+\epsilon)\{1-(1+\omega)\beta\}
\end{eqnarray*}
\begin{equation}
-\frac{\alpha}{9}\left(\frac{\dot{a}}{a}+2\frac{\dot{b}}{b}\right)^{2}
(1+\epsilon)\{1-(1+\omega)\beta\}=0\hspace{-1in}
\end{equation}

Also the differential equation for $\Lambda$ (i.e., eq.(14)) using
equations (15) and (17) can be solved as

$$
\Lambda=\omega\beta(1+\epsilon)\left(ab^{2}\right)^{-\omega\beta(1+\epsilon)}
\int\left(ab^{2}\right)^{\omega\beta(1+\epsilon)}\left(\frac{\dot{a}}{a}+2\frac{\dot{b}}{b}\right)
\left(\frac{\dot{b^{2}}}{b^{2}}
+2\frac{\dot{a}}{a}\frac{\dot{b}}{b}+\frac{k}{b^{2}}\right)dt+\Lambda_{0}
$$

Now for solution let us assume a power-law form for $b$ as

\begin{equation}
b=b_{0}t^{\nu}~,~~~b_{0},~\nu~~\text{are~constants}.
\end{equation}

Then from the field equations (10) and (11) the differential
equation for `$a$' becomes

\begin{equation}
t^{2}\ddot{a}-\nu(2\nu-1)a+\nu
t\dot{a}-\frac{k}{b_{0}^{2}}t^{2(1-\nu)}a=0
\end{equation}

For $k\ne 0$, it has simple solution for $\nu=1$ and $\frac{1}{2}$
as follows:

\begin{equation}
\nu=1~:~~~~~~~~a=a_{0}t^{\sqrt{1+\frac{k}{b_{0}^{2}}}},~~b=b_{0}t\hspace{.6in}
\end{equation}
\begin{equation}
\nu=\frac{1}{2}~:~~~~~~~a=a_{0}\text{sinh}\left(\frac{2\sqrt{|k|}}{C}t^{\frac{1}{2}}\right),~~b=b_{0}t^{\frac{1}{2}}
\end{equation}

However, for general $\nu$ the solution can be written as

\begin{equation}
a=t^{\frac{1-\nu}{2}}\left\{a_{0}~I_{\frac{(1-3\nu)}{2(1-\nu)}}[\frac{\sqrt{k}}{(1-\nu)b_{0}}t^{1-\nu}]+
a_{1}~I_{-\frac{(1-3\nu)}{2(1-\nu)}}[\frac{\sqrt{k}}{(1-\nu)b_{0}}t^{1-\nu}]\right\}
\end{equation}

where $I$ is the usual modified Bessel function.\\

Note that for the solution with $\nu=1$ the constraint satisfied
by the constants appearing in equation (18) has the simple form

\begin{equation}
(1+\epsilon)\{1-(1+\omega)\beta\}\left(3-\alpha+\frac{k}{b_{0}^{2}}\right)=2
\end{equation}

Further, for $k=0$, equation (20) has the general solution

\begin{equation}
a=a_{0}t^{1-\nu}+a_{1}t^{1-2\nu}
\end{equation}

In recent past, Sahni et al [8] proposed statefinder parameters
to discriminate between different dark energy models. In FRW model
with scale factor $a(t)$, the two parameters are defined as

\begin{equation}
r=\frac{\dddot{a}}{aH^{3}}~~,~~s=\frac{r-1}{3\left(q-\frac{1}{2}\right)}
\end{equation}

with $H$ and $q$ as the Hubble parameter and deceleration
parameter respectively. These dimensionless parameters can be
written in general form for any space-time as

\begin{equation}
r=1+\frac{3\dot{H}}{H^{2}}+\frac{\ddot{H}}{H^{3}}~~,~~s=\frac{r-1}{3\left(q-\frac{1}{2}\right)}~~,
~~q=-1-\frac{\dot{H}}{H^{2}}
\end{equation}

In the recent model we shall determine these statefinder
parameters for different solutions and examine their behaviour.\\

For $\nu=1,~k\ne 0,~q,~r$ and $s$ are all constants and
$\lambda\propto t^{-2}$ with expressions:
$$
q=\frac{1-\xi}{2+\xi}~,~r=1-\frac{9\xi}{(2+\xi)^{2}}~,~s=\frac{6}{2+\xi}~,~
\xi=\sqrt{1+\frac{k}{b_{0}^{2}}}~,~\Lambda=\frac{\Lambda_{0}}{t^{2}}
$$

\begin{figure}
\includegraphics[height=1.7in]{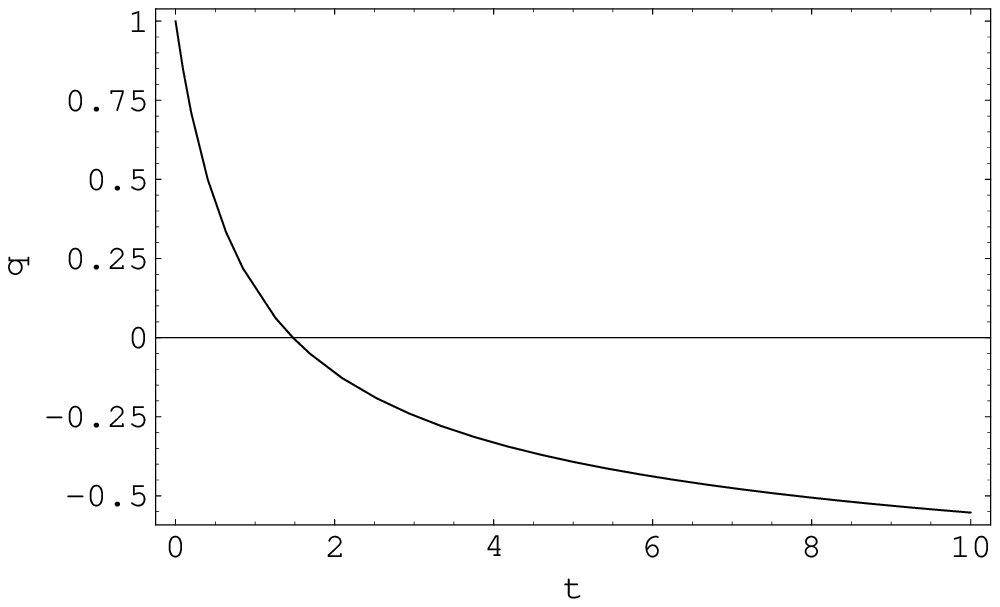}~~~
\includegraphics[height=1.7in]{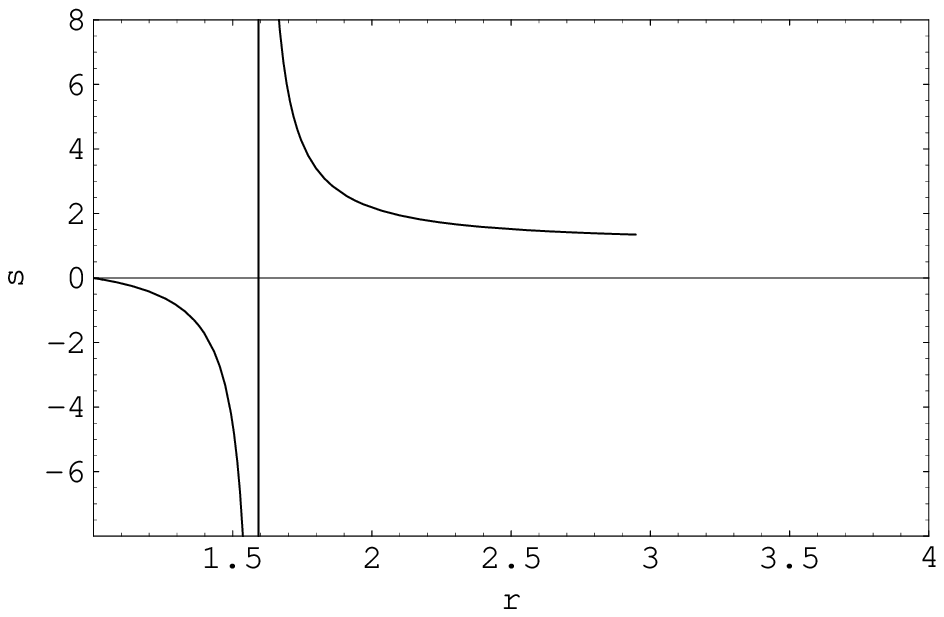}\\
\vspace{1mm}
Fig.1~~~~~~~~~~~~~~~~~~~~~~~~~~~~~~~~~~~~~~~~~~~~~~~~~~~~~Fig.2\\

\vspace{7mm} Fig.1 shows the time variation of $q$ while Fig.2
represents the variation of the parameters $r$ and $s$ for $k\ne
0$ and $\nu=\frac{1}{2}$.\hspace{10.6cm} \vspace{6mm}

\includegraphics[height=1.7in]{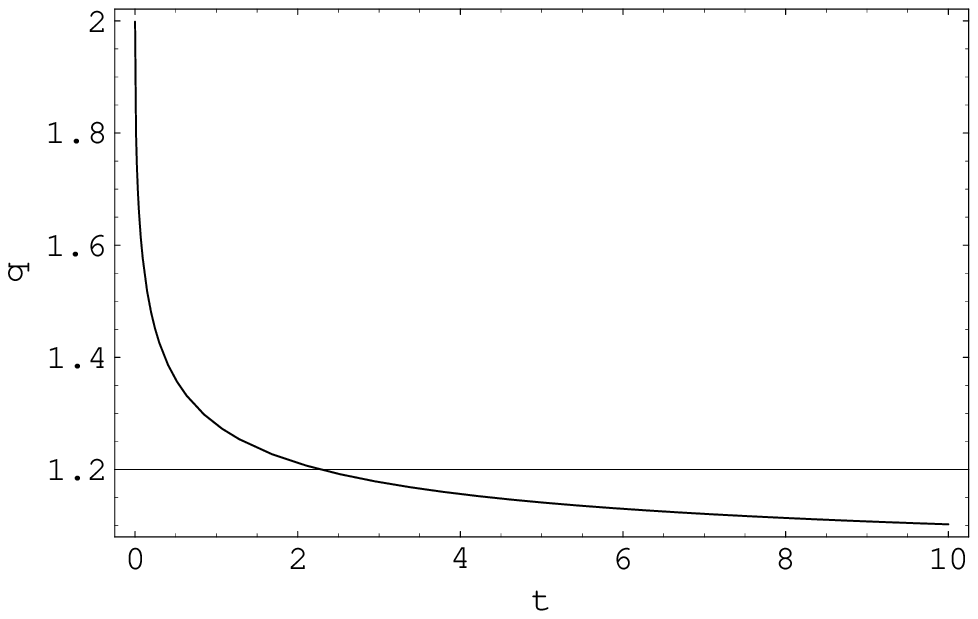}~~~
\includegraphics[height=1.7in]{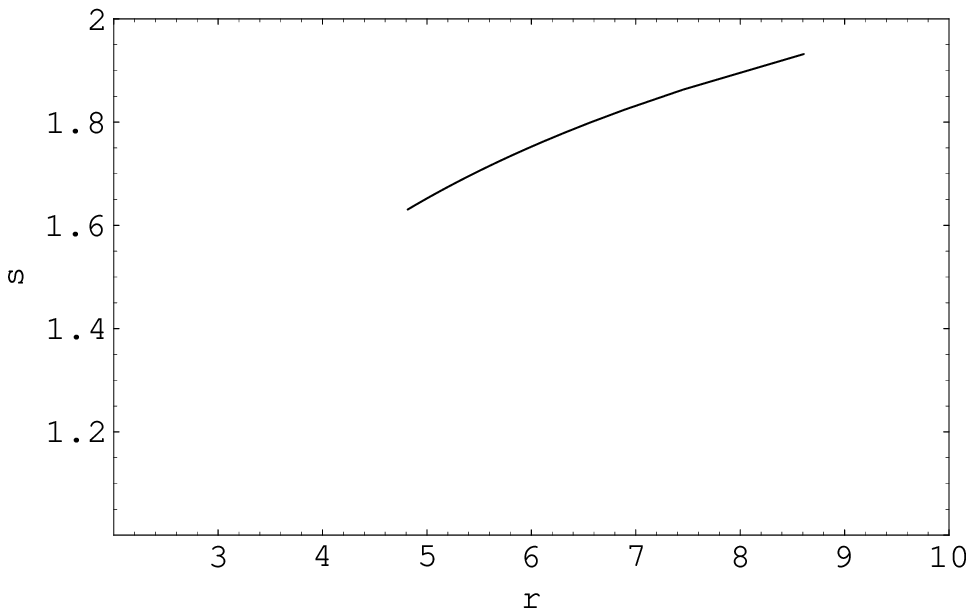}\\
\vspace{1mm}
Fig.3~~~~~~~~~~~~~~~~~~~~~~~~~~~~~~~~~~~~~~~~~~~~~~~~~~~~Fig.4\\

\vspace{5mm} Fig.3 shows the time variation of $q$ while Fig.4
represents the variation of the parameters $r$ and $s$ for $k=0$
and $\nu=\frac{1}{2}$.\hspace{10.6cm} \vspace{6mm}

\end{figure}

Note that $s$ is positive definite for all $k$ while the universe
will be accelerating for closed model and decelerating for open
model. For other $\nu$ or for flat model the expressions of these
parameters are very complicated so we have shown their behaviour
graphically only for $\nu=\frac{1}{2}$ with $k\ne 0$ or $k=0$.\\

In figures 1 and 2, we have shown the time variation of $q$ and
$r$-$s$ curve respectively for $\nu=\frac{1}{2}$, $k\ne 0$. The
figure for $q$ shows that the universe starts from an decelerating
phase (radiation era) to an accelerating phase and finally becomes
a $\Lambda$CDM model. This is also reflected in the $r$-$s$ curve.
The branch on the r.h.s. of the asymptote (in Fig.2) represents
the universe from radiation era to dust phase (when $s$ becomes
infinity) while the left portion of the curve continues from the
dust phase and
goes gradually to $\Lambda$CDM model.\\

For flat case (i.e., $k=0$) with $\nu=\frac{1}{2}$ the time
variation of $q$ and $r$-$s$ curve are shown in figures 3 and 4.
These are not of much interest from observational point of view as
the model does not represent any accelerating phase of the
universe (since $q$ is never negative).\\

Therefore, the present model with non-flat universe and
$\nu=\frac{1}{2}$ can describe the evolution of the universe from
radiation era to $\Lambda$CDM model but beyond radiation in the
past is not describable by the model $-$ probably some quantum
theory is necessary to describe the evolution of the early era.\\

{\bf Acknowledgement:}\\

One of the authors (U.D) is thankful to CSIR (Govt. of India) for
awarding a Senior Research Fellowship.\\

{\bf References:}\\
\\
$[1]$ G. Gammow, (1970) {\it My World Line}, New York.\\
$[2]$ N. R. Tanviretal, {\it Nature} London {\bf 377}, 27 (1995);
W. Freedman, astro-ph/9612204; A. G. Riess, W. H. Press and R.
Kirshner, {\it Astrophys. J.} {\bf 438}, L17 {1995}; M. Hamuyetal,
{\it Astron. J.} {\bf 112}, 2398 {1996}.\\
$[3]$ S. Perlmutter, {\it Astrophys. J.} {\bf 517} 565 (1998).\\
$[4]$ V. Silveira and I. Waga, {\it Phys. Rev. D} {\bf 56} 4625
(1997); S. Carneir  and J. A. S. Lima, gr-qc/0405141.\\
$[5]$ M. de Campos, {\it Gen. Rel. Grav.} {\bf 35} 899 (2003).\\
$[6]$ R. G. Vishwakarma, {\it Class. Quantum Grav.} {\bf 19} 4747
(2002).\\
$[7]$ M. de Campos and N. Tomimura, {\it Bazilian J. Phys.} {\bf
31} 468 (2001); J. A. Belinchon, {\it Gen. Rel. Grav.} {bf 32} 1487 (2000).\\
$[8]$ V. Sahni, T. D. Saini, A. A. Starobinsky and U. Alam, {\it
JETP Lett.} {\bf 77} 201 (2003).\\

\end{document}